\documentclass[aps, prd, twocolumn, showpacs, superscriptaddress, groupedaddress]{revtex4} 
\usepackage{graphicx}	
\usepackage{amssymb}
\usepackage{dcolumn}
\usepackage{color}
\usepackage{subfigure, rotating, bm, array}
\usepackage[pagebackref=false, colorlinks=true]{hyperref}
\hypersetup{
linkcolor=blue,     % color of internal links
citecolor=blue,     % color of links to bibliography
urlcolor=blue} 
\begin{document}
\title{Perihelion Precession and Shadows near Blackholes and Naked Singularities}
\author{Dipanjan Dey}
\email{dipanjandey.adm@charusat.edu.in}
\affiliation{International Center for Cosmology, Charusat University, Anand 388421, Gujarat, India}
\author{Rajibul Shaikh}
\email{rshaikh@iitk.ac.in}
\affiliation{Indian Institute of Technology, Kanpur, India}
\author{Pankaj S. Joshi}
\email{psjprovost@charusat.ac.in}
\affiliation{International Center for Cosmology, Charusat University, Anand 388421, Gujarat, India}

\date{\today}

\begin{abstract}
It is now known that, apart from black holes, some naked singularities can also cast shadows which provide their possible observable signatures. We examine the relevant question here as to how to distinguish then these entities from each other, in terms of further physical signatures. We point out that 
black holes always admit timelike bound orbits having positive perihelion precession. Also, while a naked singularity with a photon sphere can cast a shadow, it could also admit positive perihelion precession for such orbits, thereby mimicking a black hole. This indicates that compact objects with photon spheres (shadows) always admit positive perihelion precession of timelike bound orbits around them. On the other hand, a naked singularity without a photon sphere could admit both positive and negative perihelion precession but need not have a shadow. In this paper, we construct a spacetime configuration which has a central naked singularity but no photon sphere, and it can give both shadow and a negative perihelion precession. Our results imply that, whereas the presence of a shadow and a positive perihelion precession implies either a black hole or a naked singularity, the presence of a shadow and a negative perihelion precession simultaneously would imply a naked singularity only. We discuss our results in the context of stellar motions (motion of the `S' stars) around the Sgr-A* galactic center.
\end{abstract}

\maketitle

\section{Introduction}
Recent observation of the shadow of M87 galactic center by the Event Horizon Telescope (EHT) collaboration \cite{Akiyama:2019fyp}, triggers a lot of attention to understand the nature and dynamics of the object at the galactic center \cite{Shaikh:2019hbm,Gralla:2019xty,Abdikamalov:2019ztb,Yan:2019etp,Vagnozzi:2019apd,Gyulchev:2019tvk,Shaikh:2019fpu}. There are a lot of literature where timelike, lightlike geodesics around the black hole and naked singularity are investigated \cite{Dey:2013yga,Dey+15,levin1,Glampedakis:2002ya, Chu:2017gvt, Dokuchaev:2015zxe,Borka:2012tj,Martinez:2019nor,Fujita:2009bp,Wang:2019rvq,Suzuki:1997by,Zhang:2018eau,Pugliese:2013zma,Farina:1993xw,Dasgupta:2012zf,Shoom:2015slu,Eva, Eva1, Eva2, tsirulev, Bambhaniya:2019pbr, Joshi:2019rdo}. Generally, shadow is considered to be formed due to the existence of a photon sphere outside the event horizon of a black hole. However, in \cite{Shaikh:2018lcc}, it is shown that a naked singularity spacetime known as Joshi-Malafarina-Narayan (JMN) spacetime \cite{JMN11} can cast similar type of shadow which is expected to be seen in a black hole spacetime. In \cite{Shaikh:2018lcc}, it is shown that only the first type of JMN spacetime (JMN1) can cast shadow with a specific range of parameter's value. JMN1 spacetime is a spherically symmetric, naked singularity spacetime which can be formed as an end state of gravitational collapse in a large comoving time \cite{JMN11}. JMN1 spacetime can be written as,
\begin{equation}
ds^2=-(1-\chi)\left(\frac{r}{r_b}\right)^{\frac{\chi}{1-\chi}}dt^2+\frac{dr^2}{1-\chi}+r^2d\Omega^2\,\, ,
\label{JMNspt}
\end{equation}
where $\chi$ is a constant parameter which can have values from zero to one and at $r=r_b$ this spacetime can be matched with external Schwarzschild spacetime. Throughout the paper, we consider Newton's gravitational constant $G_N=1$ and light velocity $C=1$. In \cite{Shaikh:2018lcc}, it is shown that for $\chi>\frac23$, JMN1 spacetime casts similar shadow as what can be seen in Schwarzschild spacetime. For both the Schwarzschild and JMN1 spacetimes, the central shadow depends upon the total Schwarzschild mass ($M_{TOT}$). 
In \cite{Shaikh:2018lcc}, the theoretical results are not compared with any observational results. The main goal of that paper is to show theoretically how a naked singularity can cast similar shadow that a black hole can cast. 

The EHT collaboration will possibly release the picture of the central supermassive object (Sgr-A*) of the Milky way in this year. On the other hand, GRAVITY, SINFONI collaborations are continuously observing the stellar motion around Sgr-A* \cite{M87,Eisenhauer:2005cv,center1}. There are many `S' stars (e.g. S02, S102, S38, etc.) which are orbiting around the Sgr-A*. Among them some stars have perihelion points very close ($\sim 0.006~ Parsec$) to the central object of our Milky way. Their orbital behaviour can revel very important information about the spacetime structure around the Sgr-A*.
In \cite{Bambh}, it is shown that in a naked singularity spacetime, the perihelion precession of the bound timelike orbits can be negative, which is never possible in a Schwarzschild black hole spacetime. We always have a positive perihelion precession in this Schwarzschild black hole case. In \cite{Dey:2019fpv}, the future trajectory of S02 star is predicted considering both positive and negative precession. The negative perihelion precession occurs when a massive particle travels less than $2\pi$ angular distance in between two successive perihelion points, whereas, for positive precession, the particle has to travel greater than $2\pi$ distance in between the two successive perihelion points.

For JMN1 spacetime, negative and positive precessions occur when $\chi<\frac13$ and $\chi>\frac13$ respectively \cite{Bambh}. Also, as shown in \cite{Shaikh:2018lcc}, the JMN1 naked singularity, when matched to an exterior Schwarzschild geometry, cast shadow for $\chi>\frac23$. Therefore, the presence of both a shadow and a positive perihelion precession can mean either a Schwarzschild black hole, or a JMN1 naked singularity which has $\chi>\frac23$ and which is matched to an exterior Schwarzschild geometry. In both these cases, the bound orbits with positive perihelion precession lie in the Schwarzschild spacetime. However, if a shadow and bound orbits with negative perihelion precession are observed simultaneously, then it cannot be explained using these two scenarios as we need either a Schwazschild black hole or a JMN1 naked singularity with $\chi>\frac23$ for the shadow and only a JMN1 naked singularity with $\chi<\frac13$ for the negative precession.

In this paper, we present a spacetime configuration which can explain this latter case where we have both shadow and negative perihelion precession simultaneously. We consider a spacetime configuration where an interior JMN1$_{int}$ with $\chi_{int}>\frac23$ is matched to a second JMN1$_{ext}$ with $\chi_{ext}<\frac23$ at some radius $r=r_{b1}$ (say) and then the second JMN1$_{ext}$ is matched to an exterior Schwarzschild geometry at some greater radius $r=r_{b2}$ (say), where $r_{b2}>r_{b1}$. As we show below, such spacetime configuration allows shadow because of JMN1$_{int}$ with $\chi_{int}>\frac23$ and timelike bound orbits with negative and positive perihelion precession, respectively, for $\chi_{ext}<\frac13$ and $\frac13<\chi_{ext}<\frac23$.
\begin{figure*}
\centering
\subfigure[$h=3$,$\gamma=0.998$]
{\includegraphics[width=80mm]{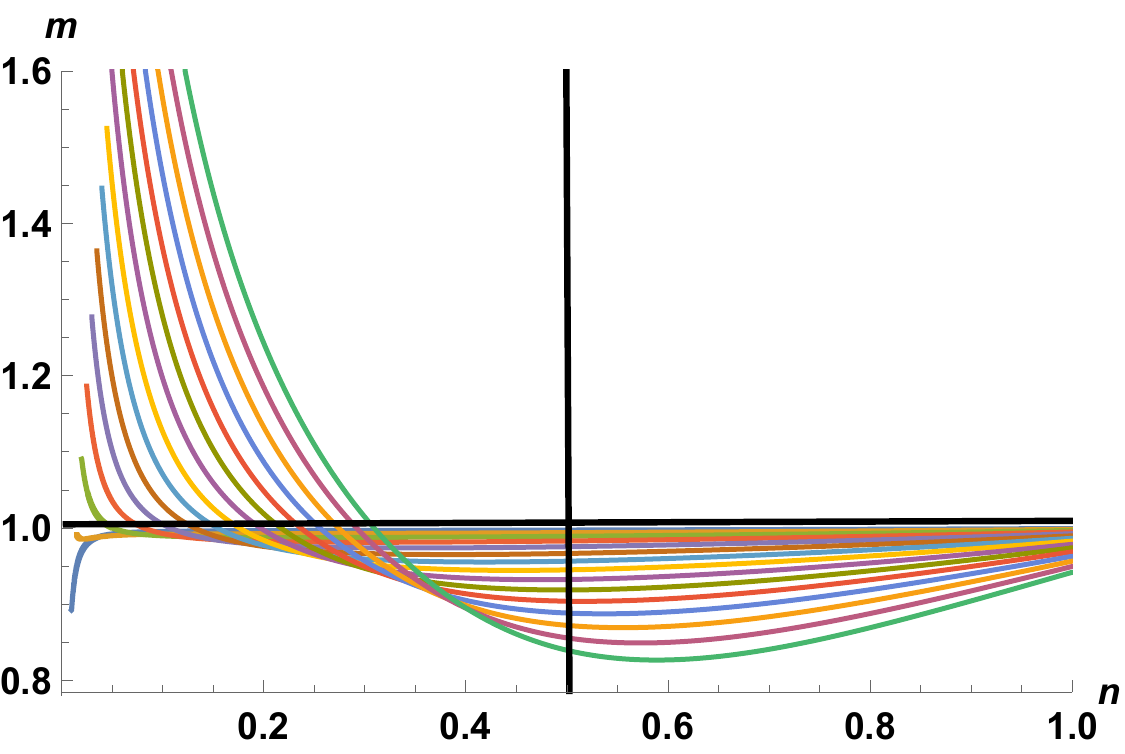}\label{re1}}
\hspace{0.1cm}
\subfigure[$h=0.1$,$\gamma=0.9797$]
{\includegraphics[width=80mm]{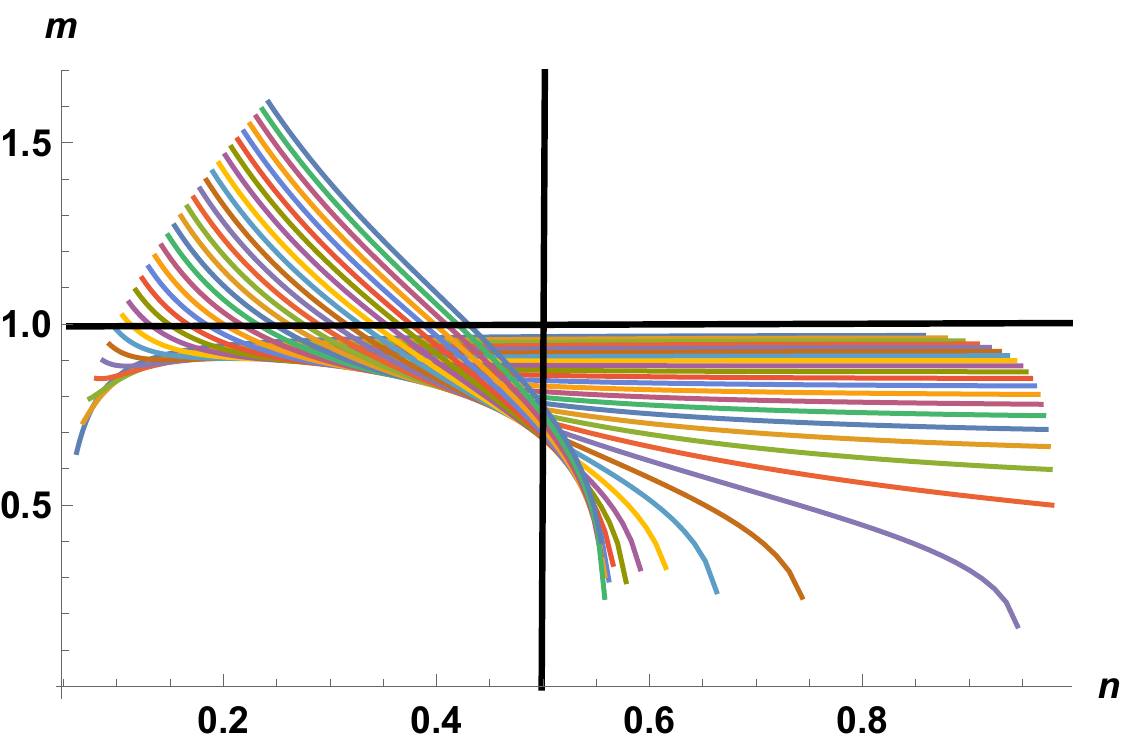}\label{re2}}
 \caption{Here, we show that the $m>1$ and $m<1$ both are possible for $n<\frac12$, however, for $n>\frac12$, only $m<1$ is possible. Along each of the different curves scalar charge $q$ varies from $0$ to $12$ and $0.06$ and $0.16$ in the fig.~(\ref{re1}) and fig.~(\ref{re2}) respectively. Each of the curves corresponds to one particular ADM mass $M$ of the JNW spacetime. In fig.~(\ref{re1}) and fig.~(\ref{re2}), $M$ varies from $0.2$ to $0.3$ and from $0.02$ to $0.03$ respectively. }
 \label{mvsn}
\end{figure*}

In the next section (\ref{structure}), we begin with a discussion of our spacetime structure. In that section, using Israel junction conditions, we show how the different $\chi$s' values can create a thin shell of matter at the matching radius. In section (\ref{shadow}), we show how a thin matter shell can create a shadow, though the photon sphere does not exist in the proposed spacetime structure, and we discuss the distinguishable properties of a black hole shadow and a shadow cast by a thin shell of matter. In that section, we also briefly review the work done in \cite{Bambh},\cite{Dey:2019fpv} and show that bound timelike orbits with positive and negative precession are possible in the JMN1$_{ext}$. In this paper, we do not compare our theoretical results with any observations. We mainly emphasize on the fact that the proposed spacetime structure with a central naked singularity allows bound timelike geodesics and a shadow of the central object which can be formed due to the presence of a thin matter shell. In section (\ref{conclusion}), we discuss our results and the outcomes implied. 
%According to the ``No-hair'' theorem a black hole can have only three properties: mass, angular momentum and charge \cite{Israel:1967wq}. Therefore, a shadow of a black hole only depends upon the above mentioned three properties of black hole. A large amount of effort is being given to understand the possible effect of extra hair on the shadow of a black hole \cite{Isi:2019aib,Loeb:2013lfa,Cunha:2015yba,Hod:2011aa}. In this paper we describe a spacetime structure which has thin matter shell at the junction of the internal and external spacetimes. We show that this spherically symmetric thin matter shell also can cast shadow whose radius depends upon the detail structure of the corresponding spacetime.

\section{Perihelion Precession and Shadow}
In \cite{Bambh}, we derive the following orbit equations for JMN1 spacetime (eq.~\ref{JMNspt}),
\begin{equation}
 \frac{d^2u}{d\phi^2} + (1 - \chi) u - \frac{\gamma^2}{2h^2}\frac{\chi}{(1- \chi)}\left(\frac{1}{u}\right)\left(\frac{1}{u~r_{b}}\right)^\frac{-\chi}{(1- \chi)}=0\,\, ,
 \label{eqJMNorbit}
\end{equation} 
where $u=\frac{1}{r}$ which is a function of azimuthal distance $\phi$. This equation can be solved numerically. In \cite{Bambh}, we solve it numerically and also present an approximate analytic solution. The approximate analytic solution of the above orbit equation can be derived by considering small value of eccentricity and it can be written as,
\begin{equation}
\tilde{u}=\frac{1}{p}\left[1+e\cos(m\phi)+O(e^2)\right]\,\, ,
\label{orbitsch1}
\end{equation}
where $\tilde{u}=ur_b$, $p$, $m$ are positive real numbers and $e$ is the eccentricity of the orbit. Using the above solution and the differential eq.~(\ref{eqJMNorbit}), one can show that for JMN spacetime the parameter $m$ can be written as,
\begin{equation}
m=\sqrt{2-3\chi}\,\, ,
\label{mJMN}
\end{equation}
which shows that for $\chi<\frac13$ and $\chi>\frac13$ we get negative $(m>1)$ and positive $(m<1)$ perihelion precession respectively. However, when we use the same approximate solution for Schwarzschild spacetime, it can be shown (\cite{Bambh}) that only positive precession of timelike bound orbit is allowed in this spacetime.
In \cite{Shaikh:2018lcc}, it is shown that, when the JMN1 spacetime with $\chi>\frac23$ is matched with an external Schwarzschild spacetime, then in the external Schwarzschild spacetime, there exist a photon sphere which casts a shadow. However, for $\chi<\frac23$, no photon sphere exist and therefore, there will be no shadow. Therefore, note that, when there is a photon sphere which cast a shadow, the perihelion precession is positive always. On the other hand, when there is no photon sphere, the perihelion precession can be both positive and negative.
%Hence, one can conclude that both shadow and negative precession of bound timelike orbit cannot exist simultaneously in that case.}

This same results exist in Janis-Newman-Winicour (JNW) naked singularity spacetime also. This spacetime is a mass-less scalar field solution of Einstein equation and it can be written as,
\begin{equation}
 ds^2_{JNW} = -\left(1-\frac{b}{r}\right)^n dt^2 + \frac{dr^2}{\left(1-\frac{b}{r}\right)^{n}} + r^2\left(1-\frac{b}{r}\right)^{1-n}d\Omega^2\,\, ,
 \label{JNWmetric}
\end{equation}
where $b=2\sqrt{M^2+q^2}$ and $n=\frac{2M}{q}$. The parameters $q$ and $M$ represent charge of the scalar field and the ADM mass respectively. From the expression of $b$ and $n$ one can show that $0<n<1$. Using eq.~(\ref{orbitsch1}) in eq. (58) of \cite{Bambh}, one can show
\begin{equation}
 m=\sqrt{Qp-2R-\frac{3S}{p}}\label{mJNW}\,\, ,
\end{equation}
where, 
$$  p_{\pm}=\frac{R\pm\sqrt{R^2+4QS}}{2Q} $$
 $$Q = \left[\frac{b^2\gamma^2(1-n)}{h^2} - \frac{b^2(2-n)}{2h^2}\right]\,\, ,$$
$$R = \left[\frac{b^2\gamma^2(1-n)(1-2n)}{h^2} - \frac{b^2(2-n)(1-n)}{2h^2} - 1\right]\,\, ,$$
$$S = \left[\frac{3}{2} - \frac{b^2(2-n)(1-n)n}{4h^2} + \frac{b^2\gamma^2(1-2n)(1-n)n}{h^2}\right]\,\, ,$$ where $\gamma$ and $h$ are the conserved energy and angular momentum per unit rest mass (see \cite{Bambh} for details). In fig.~(\ref{mvsn}), it is shown that stable bound orbit with negative precession ($m>1$) is only possible when $n<\frac12$. On the other hand, positive precession ($m<1$) of timelike bound orbit is possible for both $n<\frac12$ and $n>\frac12$. In \cite{Shaikh:2019hbm}, it is shown that for $n<\frac12$, JNW spacetime cannot cast a shadow. JNW can cast shadow only for $\frac12<n<1$. 

Therefore, from the above results, we can conclude that, when both the JMN1 and JNW naked singularities have photon spheres and hence cast shadows, they admit only positive perihelion precession of bound orbits. On the other hand, when they do not have any photon sphere, they admit both negative and positive precession but no shadows. Then the question arises is, can both shadows and negative precession exist simultaneously? In the next section,  using an internal JMN1$_{int}$ and  an external JMN1$_{ext}$, we construct a spacetime configuration which has a central naked singularity but no photon sphere, and it can give both shadow and a negative perihelion precession.

\section{The Spacetime Structure}
\label{structure} In \cite{Dey:2019fja}, we discussed how a galactic halo like structure can form due the gravitational collapse of General Collapsing Metric (GCM), where we model the GCM as the spacetime of collapsing baryonic matter and dark matter. In general relativity, a spherically symmetric general collapsing metric can be written as,

\begin{equation}
    ds^2_{\text{GCM}} = - e^{2\nu(r,t)} dt^2 + {R'^2\over G(r,t)}dr^2 + R^2(r,t) d\Omega^2\,\, ,
\end{equation}

where $r,t$ are the comoving radial and temporal coordinates respectively and $R(r,t)$ is the physical radius. In the above equation $G(r,t)$ and $\nu(r,t)$ are the functions of comoving radius and comoving time, where $G(r,t)$ can have  positive values only. In \cite{Dey:2019fja}, we considered the above metric to be seeded by baryonic matter and dark matter which are collapsing together quasistatically at the initial stage of gravitational collapse. As the cooling time of baryonic matter is less than its dynamic time, baryonic matter cools down and accumulates at the central region of halo. In \cite{Dey:2019fja}, this situation is described by an internal and an external GCM spacetimes.
% One can show that any mismatch in $G(r,t)$ of the internal and external GCM would cause a thin matter shell which would collapse along with the whole dynamic spacetime structure.
 It can be shown that a collapsing matter cloud can reach to an equilibrium state after a large enough comoving time if there exists a non-zero pressure\cite{JMN11}. Similarly, the spacetime structure made with an internal GCM and an external GCM can also reach to an equilibrium state in asymptotic time. In \cite{Dey:2019fja}, it is shown that the above mentioned spacetime structure, in a large comoving time, can transform into a static, spherically symmetric spacetime structure which can be described by an internal JMN1 (JMN1$_{int}$)and an external JMN1 (JMN1$_{ext}$) spacetime, where JMN1$_{ext}$ is matched with an external Schwarzschild spacetime. This static spacetime structure can be described as, 
\begin{figure*}[t!]
%\centering
\subfigure[Effective ptential of lightlike geodesic in Schwarzschild spacetime with $M_{TOT}=0.5$]
{\includegraphics[width=60mm]{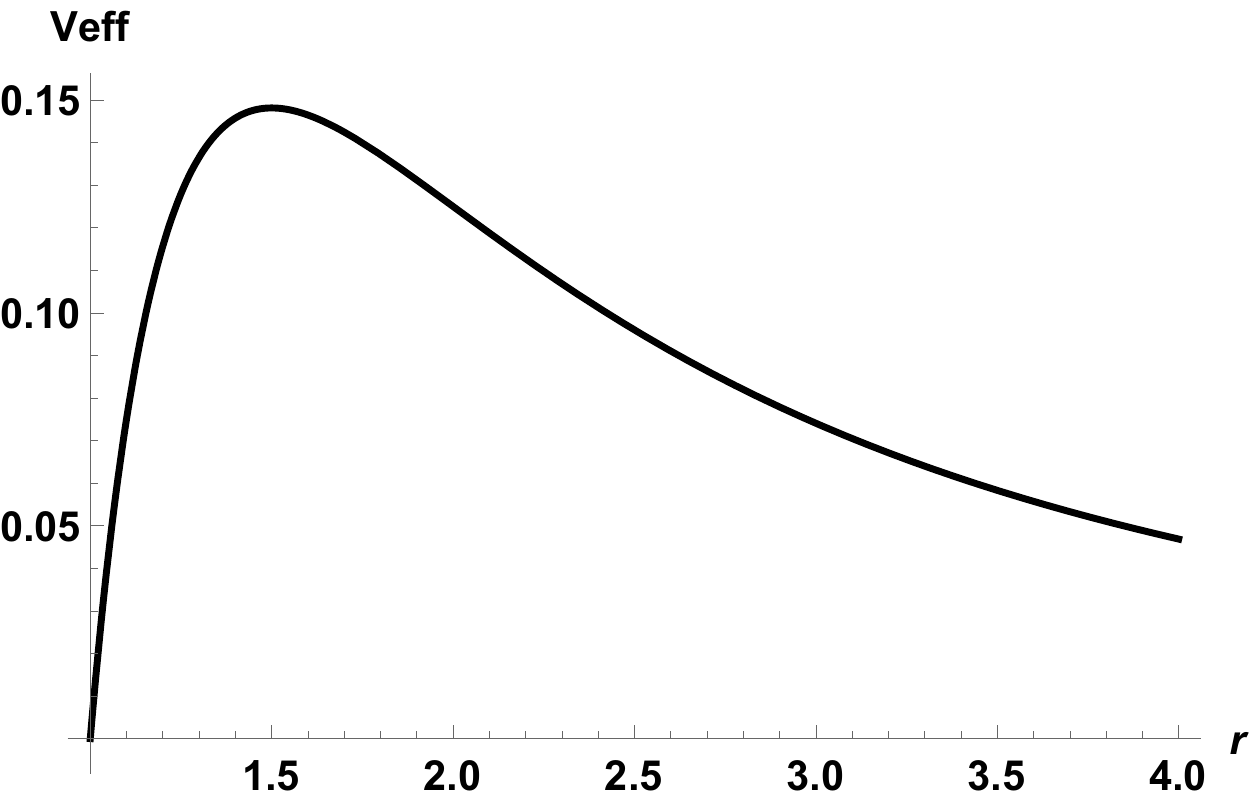}\label{Schpoten}}
\subfigure[Intensity variation along the X-axis for Schwarzschild spacetime (with $M_{TOT}=0.5$) in observer sky]
{\includegraphics[width=55mm]{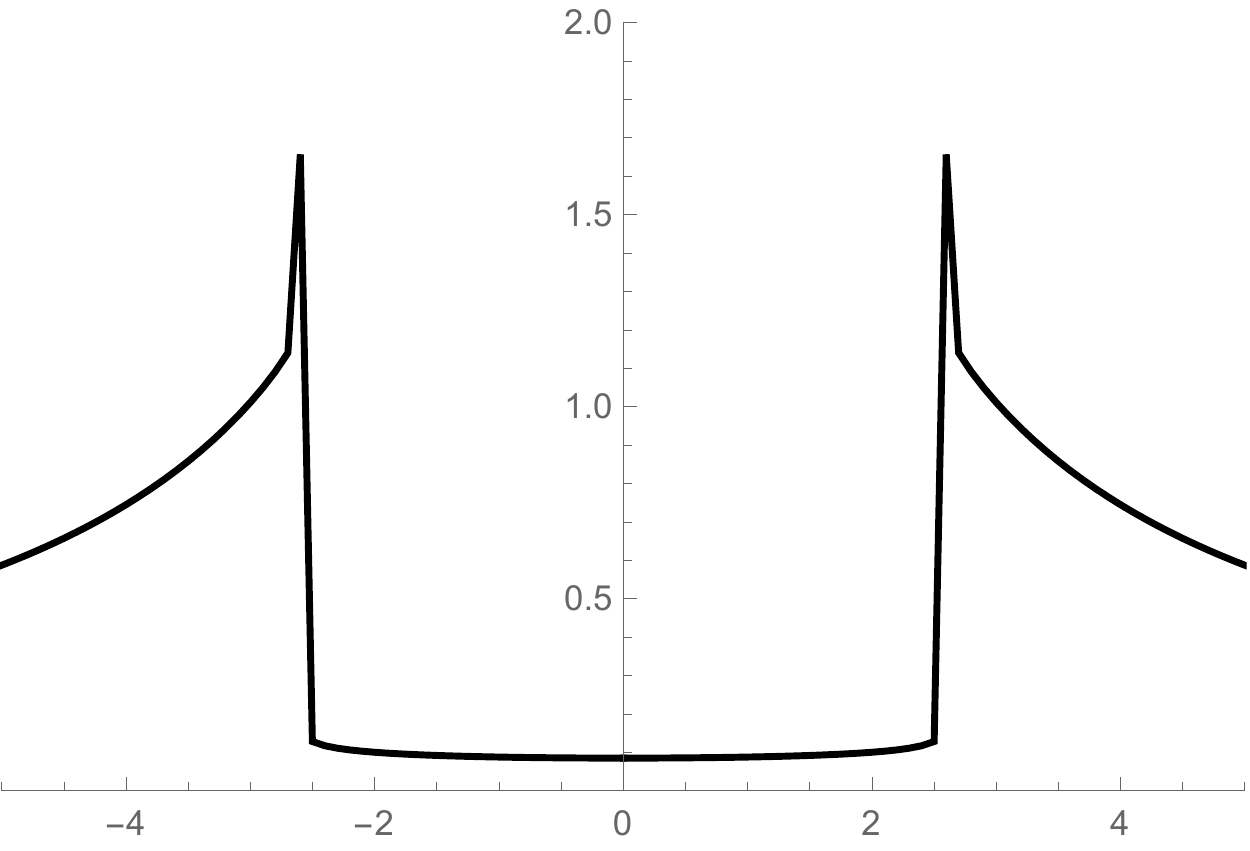}\label{schinten}}
\subfigure[Shadow of the photon sphere in Schwarzschild spacetime with $M_{TOT}=0.5$]
{\includegraphics[width=55mm]{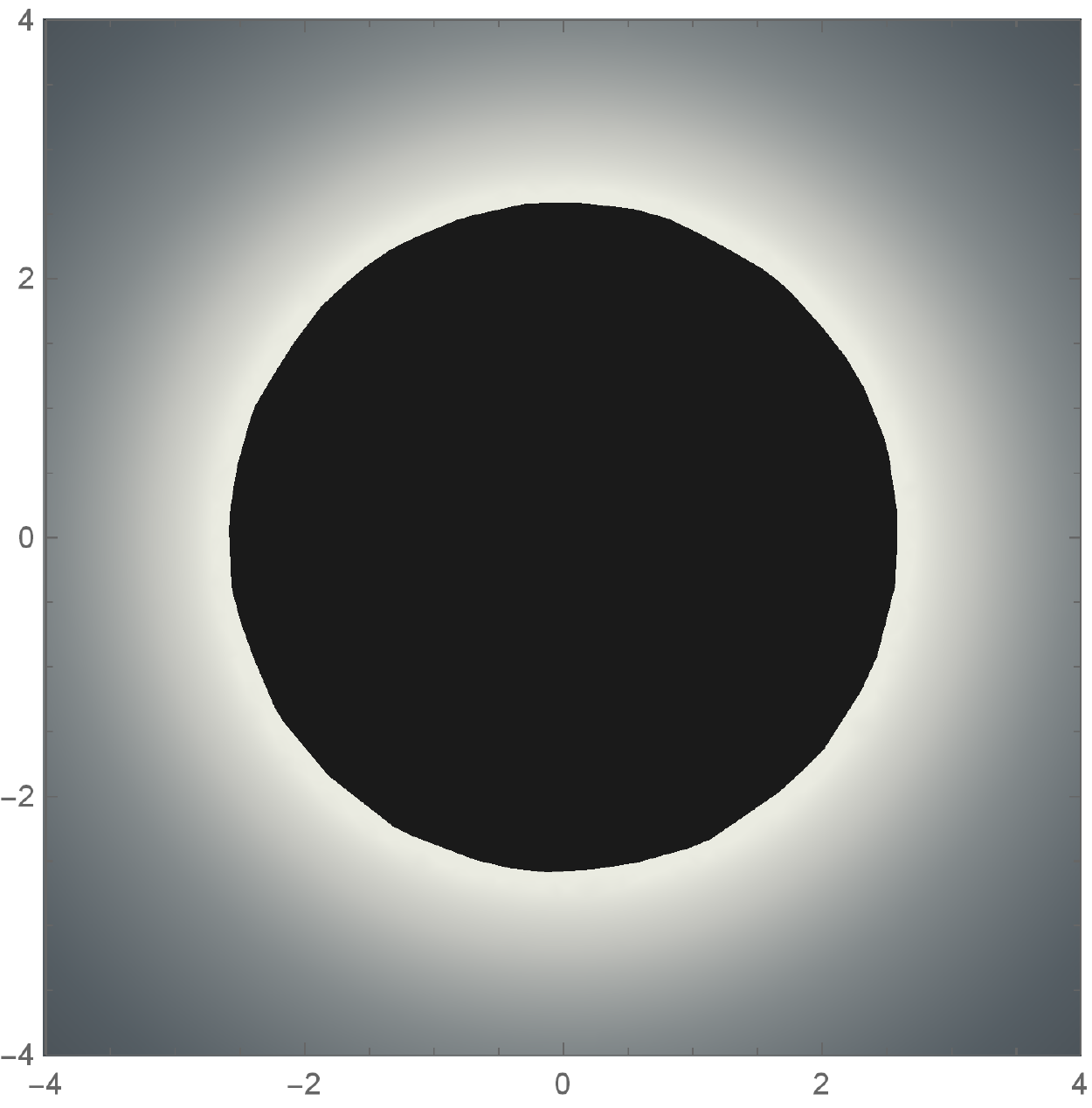}\label{schshadow}}
\subfigure[Effective ptential of lightlike geodesic in a spacetime structure]
{\includegraphics[width=60mm]{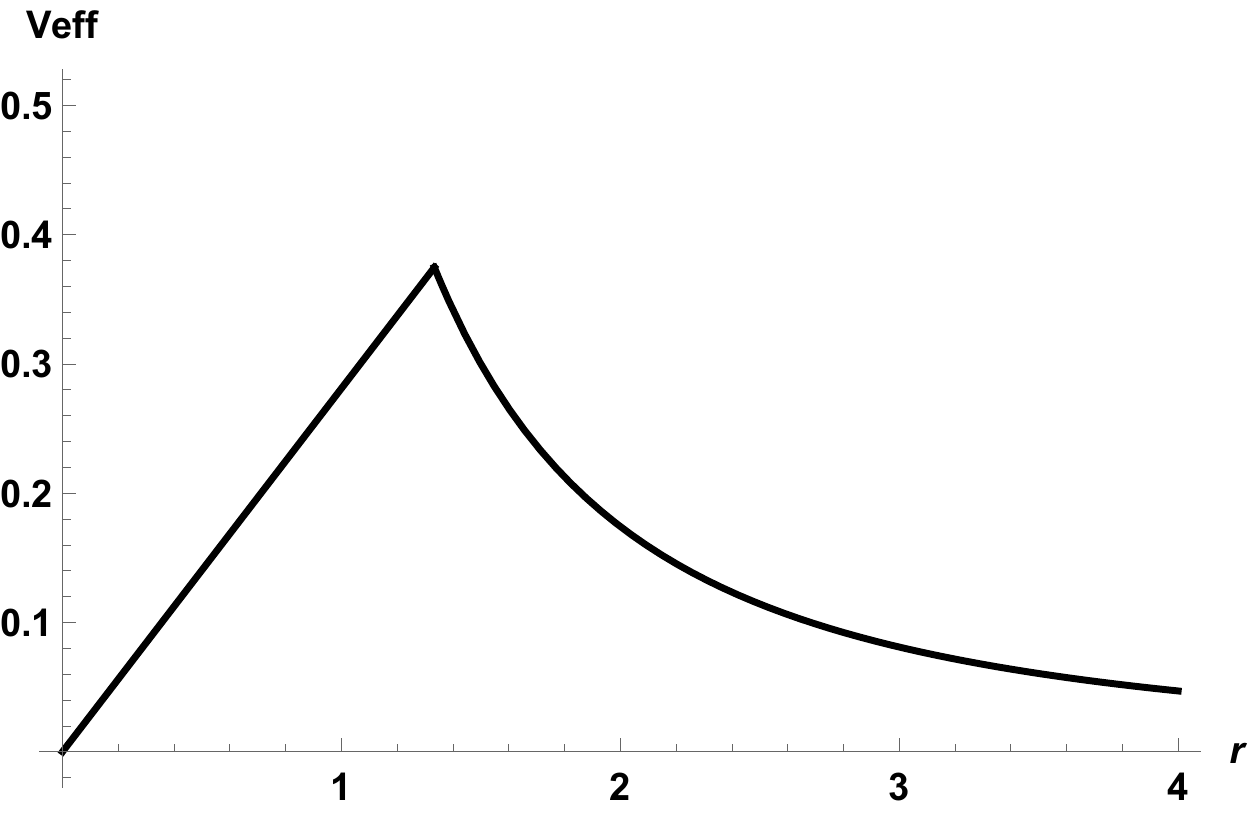}\label{JMNpoten}}
\subfigure[Intensity variation along the X-axis in the observer sky for the spacetime structure mentioned in text.]
{\includegraphics[width=55mm]{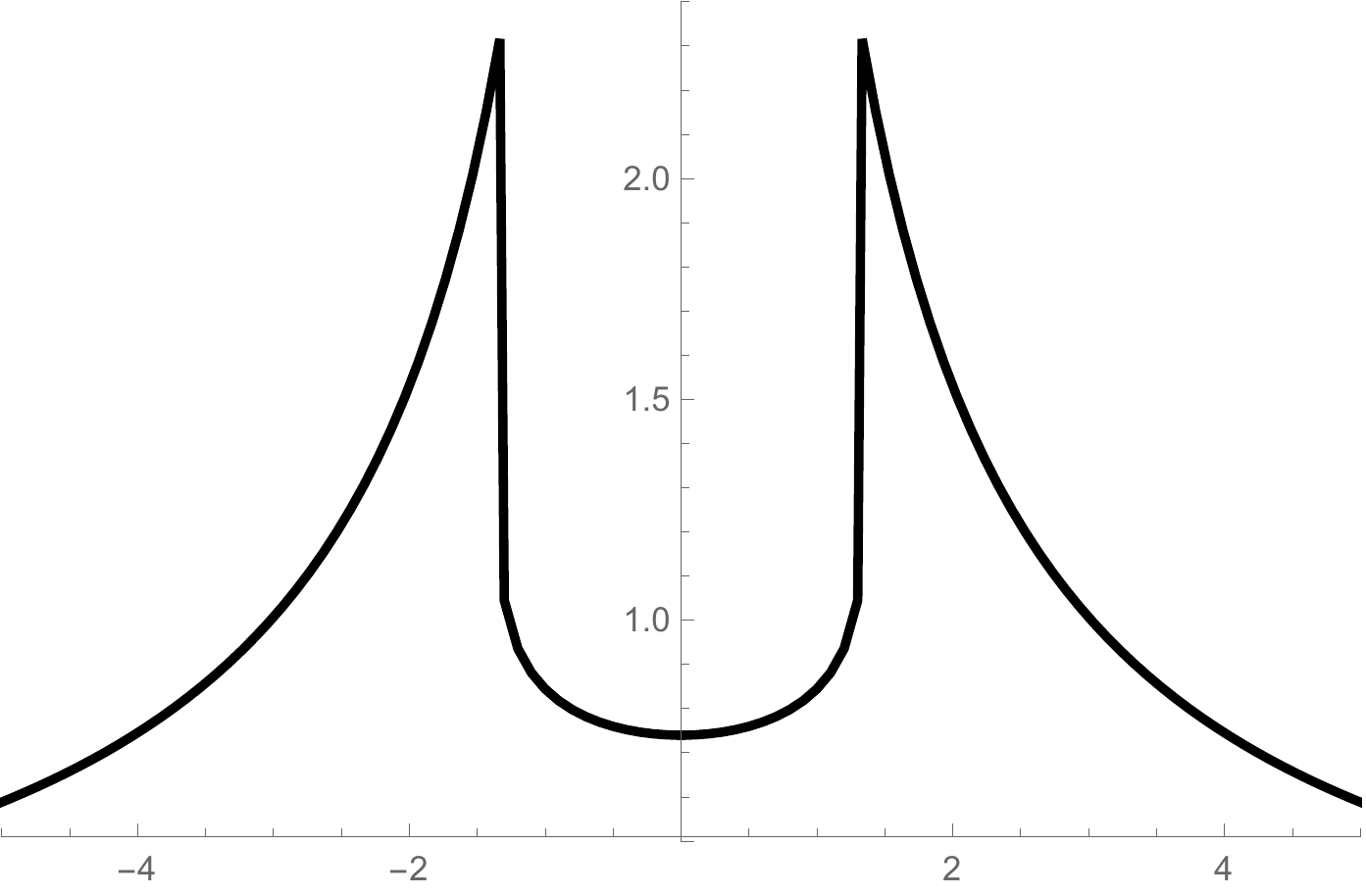}\label{JMNinten}}
\subfigure[Shadow of a spherically symmetric, thin matter shell.]
{\includegraphics[width=55mm]{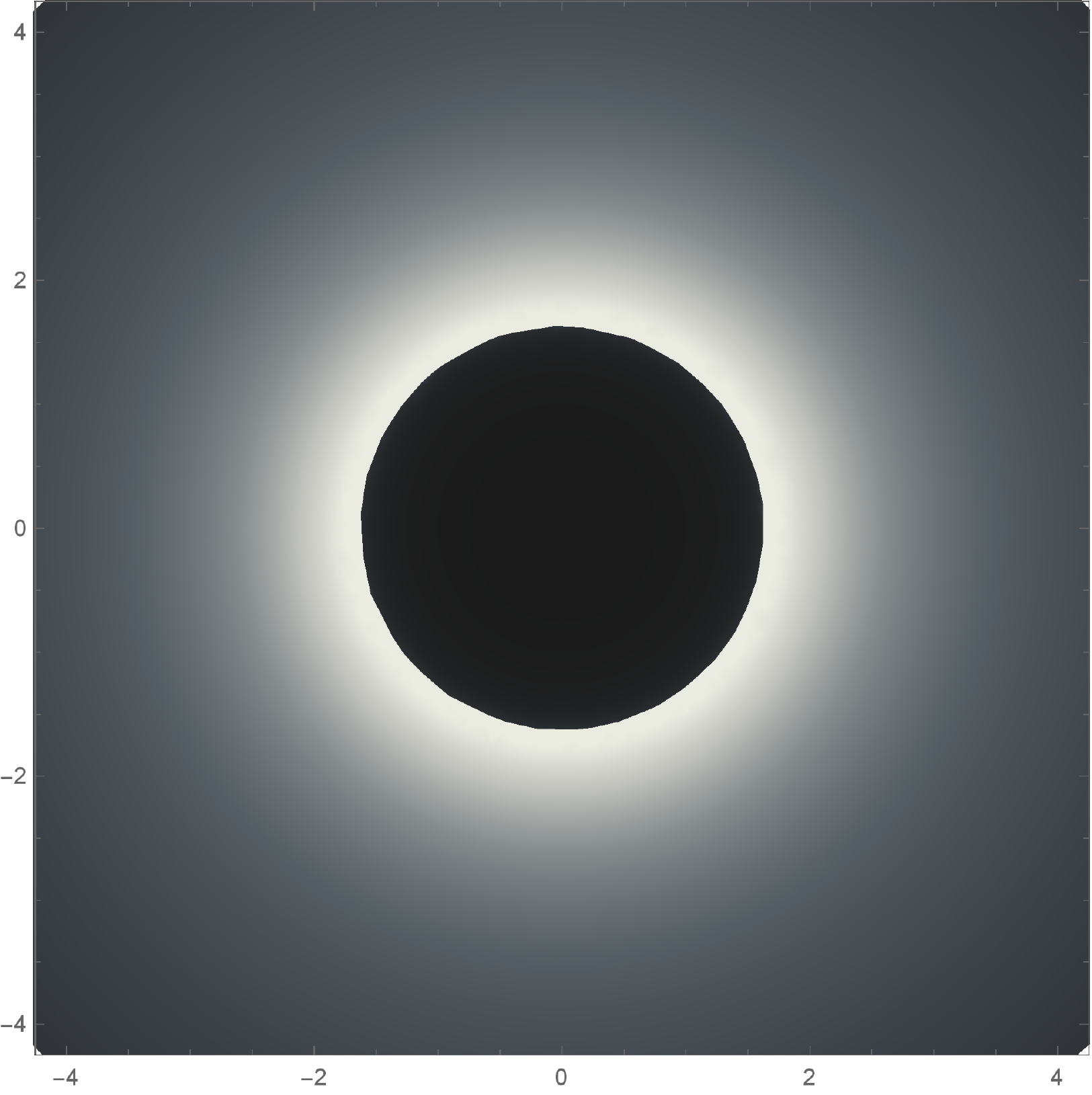}\label{JMNshadow}}
 \caption{ In this figure, the effective potential ($V_{eff}$) for null geodesics, the intensity map in observer sky and the shadow of the central object are shown for Schwarzschild spacetime and the spacetime structure which has an internal and an external JMN1 spacetimes. The shadow shown in right bottom corner is the shadow of a thin matter shell which is formed due to the mismatch in $\chi$ of internal and external JMN1 spacetime. Here we take $\chi_{int}=0.75$ and $\chi_{ext}=0.1$ for JMN1$_{int}$ and JMN1$_{ext}$ spacetimes respectively.}
\label{JMNSCHshadow}
\end{figure*}
\begin{widetext}
\begin{eqnarray}
ds^2_{\text{int}} &=& -(1-\chi_{ext})\left(\frac{\tilde{r}_{b1}}{\tilde{r}_{b2}}\right)^{\frac{\chi_{ext}}{1-\chi_{ext}}}\left(\frac{r}{r_{b1}}\right)^{\frac{\chi_{int}}{1-\chi_{int}}}dt^2 \nonumber + \frac{dr^2}{1-\chi_{int}} + r^2 d\Omega^2, \\
ds^2_{\text{ext}} &=& -(1-\chi_{ext})\left(\frac{\tilde{r}}{\tilde{r}_{b2}}\right)^{\frac{\chi_{ext}}{1-\chi_{ext}}}dt^2 \! + \! \frac{d\tilde{r}^2}{1-\chi_{ext}} \! + \! \tilde{r}^2 d\Omega^2. \nonumber\\
ds^2_{\text{Schw}}&=&-\left(1 - \frac{\chi_{ext} \tilde{r}_{b2}}{\tilde{r}}\right)dt^2 \! + \! \frac{d\tilde{r}^2}{\left(1 - \frac{\chi_{ext} \tilde{r}_{b2}}{\tilde{r}}\right)} \! + \! \tilde{r}^2d\Omega^2\,\, ,
\label{spctimestruc}
\end{eqnarray}
\end{widetext}
where JMN1$_{int}$ and JMN1$_{ext}$ are matched at a timelike hypersurface $r-r_{b1}=0$ and JMN1$_{ext}$ is matched with external Schwarzschild metric at a timelike hypersurface $r-r_{b2}=0$.
For the smooth matching of JMN1$_{int}$ and JMN1$_{ext}$, we need to match the induced metrics ($h_{ab}$) and the extrinsic curvatures ($K_{ab}$) at the matching hypersurface $r-r_{b1}=0$. It can be shown that, in order to match the induced metrics and extrinsic curvatures for JMN1$_{int}$ and JMN1$_{ext}$, we need $\tilde{r}=r$ and $\chi_{int}=\chi_{ext}$. Since the spacetime configuration we are considering has a mismatch in $\chi$ of internal and external asymptotic spacetimes (JMN1$_{int}$ and JMN1$_{ext}$), then there is a spherically symmetric thin matter shell at the timelike hypersurface $r-r_{b1}=0$. The external JMN1$_{ext}$, however, is smoothely matched to the external Schwarzschild spacetime \cite{Shaikh:2018lcc}. In the next section, we discuss shadows and prehelion precession in the above-mentioned spacetime.

\section{Shadow cast by the proposed spacetime configuration }
\label{shadow}
%\section{Shadow of thin matter shell}
A spherically symmetric static spacetimes can be written as,
\begin{equation}
ds^2=-A(r)dt^2+B(r)dr^2+r^2d\Omega^2\,\, ,
\label{sptm}
\end{equation}
where $A(r)$ and $B(r)$ are the positive valued functions of $r$. For light like geodesics in these spacetimes, we can write,
\begin{equation}
A(r)B(r)\left(\frac{dr}{d\lambda}\right)^2+V_{eff}=e^2\,\, ,
\label{sphersptm}
\end{equation}
where $\lambda$ is the affine parameter and $V_{eff}$ is effective potential of lightlike geodesics which can be written as, $V_{eff}=\frac{A(r)}{B(r)}l^2$, where $e$ and $l$ are the conserved energy and angular momentum of photon. In the above equation, we use $k_{\mu}k^{\mu}=0$, where $k^{\mu}$ is the nulllike four velocity. From the effective potential of photon, one can get the information about the turning points and stable and unstable circular orbits of light like geodesics. When the effective potential has a maximum point at $r_{ph}$, where $V_{eff}(r_{ph})=e^2$, $V^{\prime}_{eff}(r_{ph})=0$ and $V^{\prime\prime}_{eff}(r_{ph})<0$, we can say that at $r=r_{ph}$, lightlike unstable, circular geodesics are possible. This timelike spherical surface of radius $r_{ph}$ is known as photon sphere. In a spherically symmetric spacetime, a photon sphere exists when above mentioned conditions of $V_{eff}$ are fulfilled. Turning points ($r_{tp}$) of lightlike geodesics can be found from  $V_{eff}(r_{tp})=e^2$. From $V_{eff}(r_{tp})=e^2$, one can write $r_{tp}=b\sqrt{A(r)}$, where $b=\frac{l}{e}$ which is known as impact parameter. 
%When the effective potential has a cusp like nature at a point $r=r_{csp}$ or $V_{eff}(r_{csp})=e^2$ and $V^{\prime\prime}_{eff}(r_{csp})\rightarrow -\infty$, light like geodesic can not have unstable circular orbits at $r=r_{csp}$. Therefore, due to the existence of cusp like potential, the lightlike geodesics can show some distinguishable properties which can not be seen when photon sphere exists. 

One can verify that photon sphere in JMN1 spacetime is not possible, as the effective potential of null geodesic cannot fulfil the previously mentioned conditions. On the other hand, in Schwarzschild spacetime photon sphere exists. In fig.~(\ref{Schpoten}), the effective potential for light like geodesics in Schwarzschild spacetime is shown, where the effective potential has a maximum point. Therefore, in Schwarzschild spacetime photon sphere exists. However, in \cite{Shaikh:2018lcc}, it is shown that a spacetime structure, which is internally JMN1 spacetime and externally Schwarzschild spacetime, can have a photon sphere in the external Schwarzschild spacetime when $\chi>\frac23$. Therefore, the spacetime structure can cast same type of shadow what a solely Schwarzschild spacetime can cast.
%%%%%%%%%%%%%%%%%%%%%%%%%%%%%%%%%%%%%%%%%
\begin{figure*}[t!]
%\centering
\subfigure[$\chi_{int}=0.75$, $\chi_{ext}=0.3$]
{\includegraphics[width=95mm]{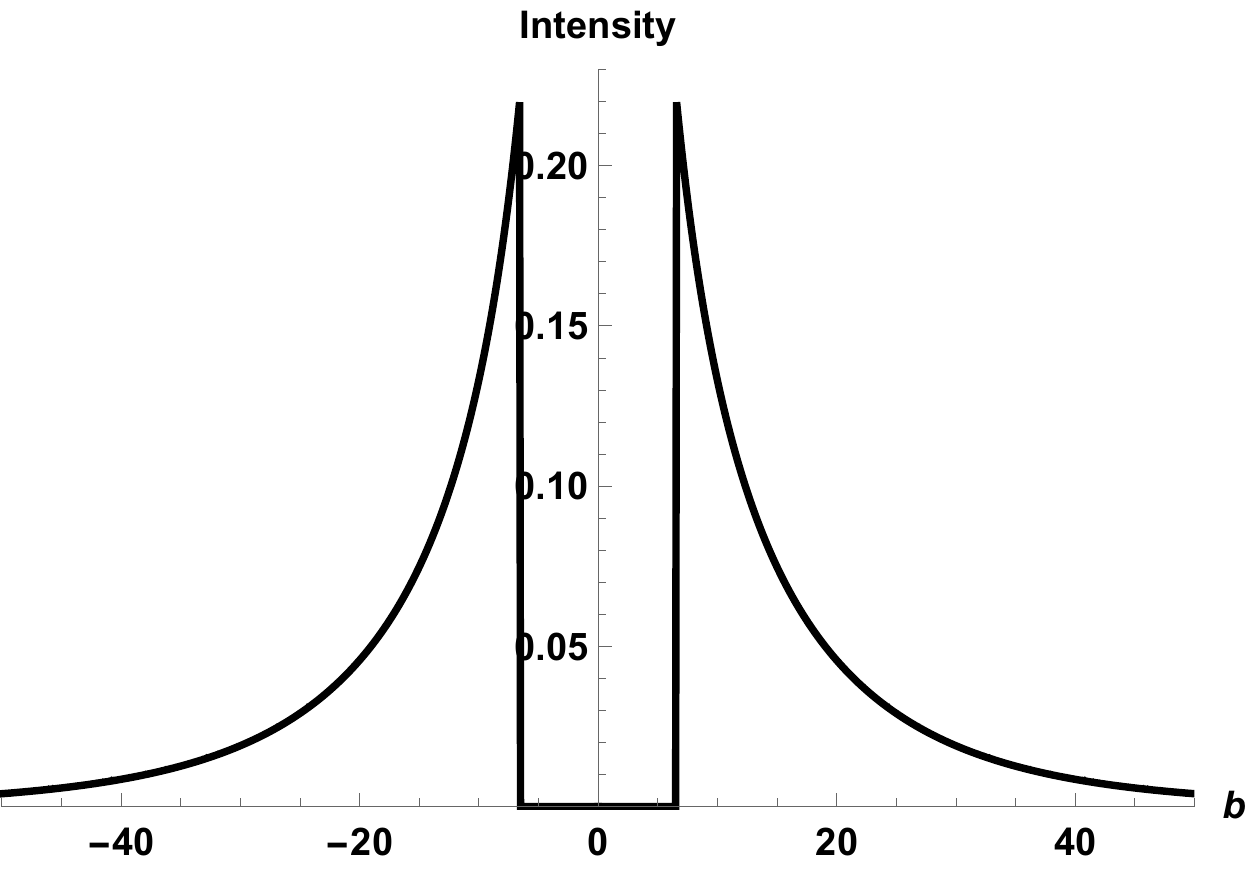}\label{JMNintensity1}}
\subfigure[$\chi_{int}=0.75$, $\chi_{ext}=0.3$]
{\includegraphics[width=80mm]{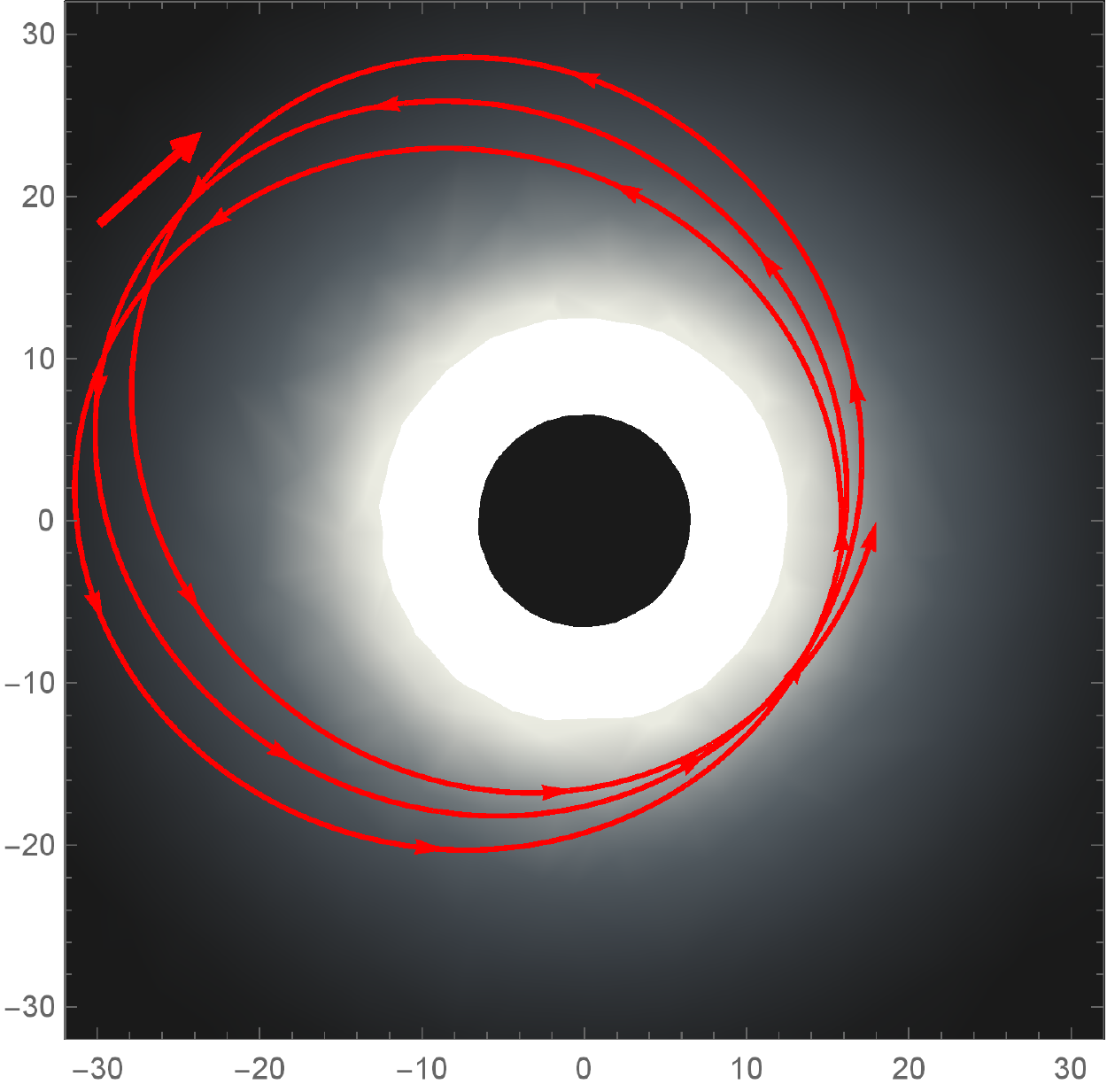}\label{JMNorbitshadow1}}\\
\subfigure[$\chi_{int}=0.75$, $\chi_{ext}=0.36$.]
{\includegraphics[width=95mm]{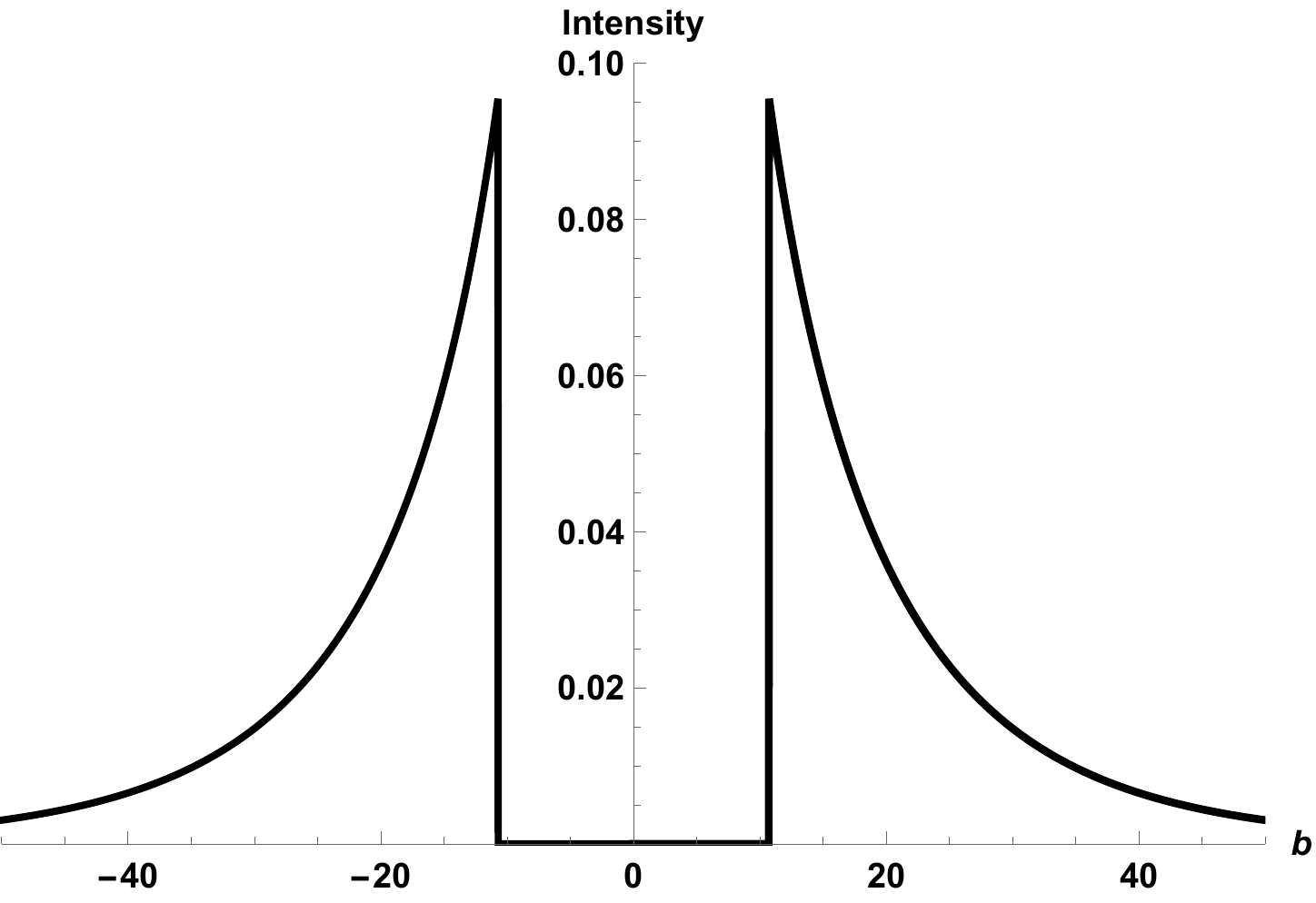}\label{JMNintensity2}}
\subfigure[$\chi_{int}=0.75$, $\chi_{ext}=0.36$.]
{\includegraphics[width=80mm]{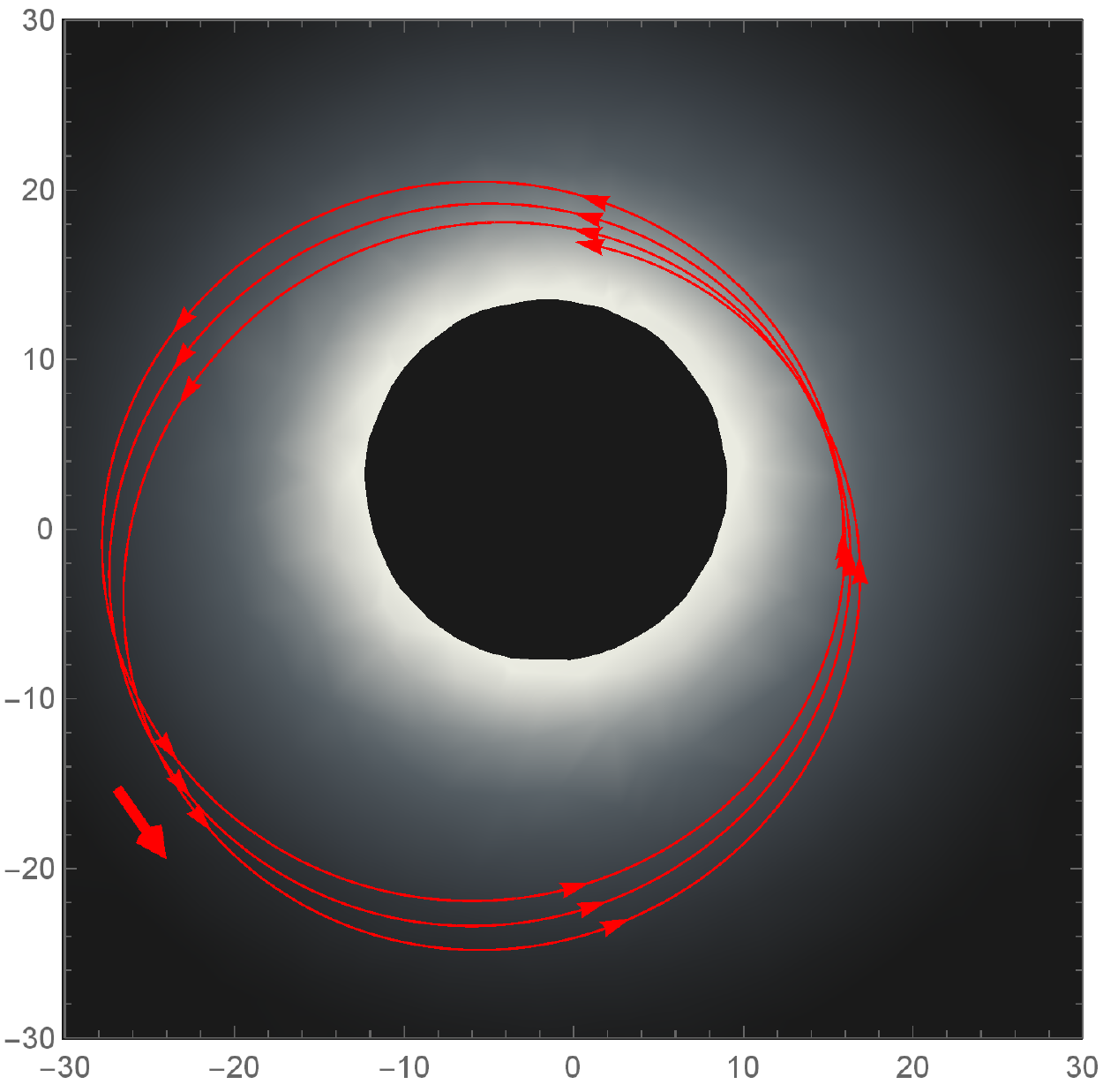}\label{JMNorbitshadow2}}
 \caption{Fig.~(\ref{JMNintensity1}) and fig.~(\ref{JMNintensity2}) show how intensity varies with the impact parameter ($b$) for $\chi_{ext}=0.3$ and $\chi_{ext}=0.36$ respectively. In fig.~(\ref{JMNorbitshadow1}), shadow with negative precession of stable timelike bound orbit (solid red line) is shown. On the other hand, fig.~(\ref{JMNorbitshadow2}) shows the existence of shadow with positive precession of stable timelike bound orbit (solid red line).}
\label{JMNSCH}
\end{figure*} 
%%%%%%%%%%%%%%%%%%%%%%%%%%%%%%%%%%%%%%%%
The spacetime structure which is mentioned in eq.~(\ref{spctimestruc}) does not allow any photon sphere inside the JMN1$_{int}$ and JMN1$_{ext}$ spacetimes. As it was discussed before,  at the junction of two JMN1 spacetimes a thin matter shell can exist due to the mismatch of $\chi_{int}$ and $\chi_{ext}$. Now, if one consider $\chi_{int}>\frac23$ and $\chi_{ext}<\frac23$ then the spacetime structure described in eq.~(\ref{spctimestruc}) have a cusp like potential as shown in fig.~(\ref{JMNpoten}). 
When the effective potential has a cusp like nature at a point $r=r_{csp}$, where $V_{eff}(r_{csp})=e^2$ and $V^{\prime\prime}_{eff}(r_{csp})\rightarrow -\infty$, light like geodesic cannot have unstable circular orbits at $r=r_{csp}$. Therefore, due to the existence of cusp like potential, the lightlike geodesics can show some distinguishable properties which cannot be seen when photon sphere exists. At the cusp point ($r_{csp}$), as the $V^{\prime}_{eff}(r)$ is non-zero, there will be no photon sphere at this point. Therefore, relativistic Einstein rings does not form due to the cusp point of the effective potential. However, incoming light like geodesics can have innermost turning point at $r=r_{csp}$. Therefore, there exist a critical impact parameter, $b_{csp}=\frac{r_{csp}}{\sqrt{A(r_{csp})}}$, corresponding to $r_{csp}$. Ingoing photons with an impact parameter greater than $b_{csp}$ must be deflected away by the cusp potential and those with impact parameter less than $b_{csp}$ would fall into the central singularity. Therefore, in such a case, we have a shadow of radius $b_{csp}$. Next, following \cite{Shaikh:2018lcc}, we consider an optically thin, radiating, radially infalling, accreting matter around the central singularity and produce the intensity map of the image. Beside the ingoing photons with impact parameter $b<b_{csp}$ getting absorbed by the central singularity, the outdoing photons which are emitted from the region $r<r_{csp}$ and have $b<b_{csp}$ can escape and are highly redshifted. Consequently, there should be a sudden drop in the observed intensity ($I_{obs}(X,Y)$) in the region $0\leq b\leq b_{csp}$, where $(X,Y)$ is the point in observer sky and impact parameter $b = \sqrt{X^2+Y^2}$. The observed intensity can be written as \cite{Bambi:2013nla} (see also \cite{Shaikh:2018lcc}),
\begin{equation}
I_{obs}(X,Y)=-\int_{\gamma}\frac{g^3k_t}{r^2 k^r}dr\,\, ,
\label{Iobs}
\end{equation}
where the integration is done along the photon trajectory ($\gamma$) and $g$, $k^t$ and $k^r$ are the redshift factor, temporal part and radial part of null four velocity respectively. The redshift factor $g$ can be written as \cite{Shaikh:2018lcc}, $$g=\frac{1}{\frac{1}{A(r)}\pm\sqrt{\left(\frac{1}{A(r)}-1\right)\left(\frac{1}{A(r)}-\frac{b^2}{r^2}\right)}}\,\, ,$$ where we consider photon trajectory inside the spherically symmetric spacetime mentioned in eq.~(\ref{sptm}). Here we consider only a simple model where a optically thin radiating matter radially freely falling towards the center with an emissivity proportional to $r^{-2}$ and the emitted radiation is monochromatic \cite{Bambi:2013nla}. We can use eq.~(\ref{Iobs}) to derive the intensity map with respect to the stationary asymptotic observer. The intensity variation and the intensity map of the images for the Schwarzschild black hole and the proposed spacetime structure
 shown in figs.~(\ref{schinten}), (\ref{schshadow}), (\ref{JMNinten}) and (\ref{JMNshadow}). In Schwarzschild black hole spacetime, it is the shadow of the photon sphere which can be observed by asymptotic observer. The photon sphere exists in Schwarzschild spacetime at $r_{ph}=3M_{TOT}$ and therefore, the radius of the shadow in the observer sky will be,
\begin{equation}
b_{ph}=\frac{r_{ph}}{\sqrt{\left(1-\frac{2M_{TOT}}{r_{ph}}\right)}}=3\sqrt3 M_{TOT}\,\, .
\label{shadowSCH}
\end{equation} 
In fig.~(\ref{JMNSCHshadow}), we consider the total mass $M_{TOT}=0.5$ for the Schwarzschild black hole. Therefore, the shadow radius will be $b_{ph}=2.59$, which can be seen in fig.~(\ref{schshadow}). On the other hand, in the spacetime structure (eq.~(\ref{spctimestruc})), it is the shadow of thin matter shell. The thin matter shell exists at a timelike hypersurface $r-r_{b1}=0$. Therefore, the shadow radius in the observer sky will be,
\begin{equation}
b_{csp}=\frac{r_{b1}}{\sqrt{1-\chi_{ext}}\left(\frac{r_{b1}}{r_{b2}}\right)^{\frac{\chi_{ext}}{2(1-\chi_{ext})}}}\,\, , 
\label{shadowStruc}
\end{equation}
where we can see that the radius of the shadow not only depends upon the radius ($r_{b1}$) of thin matter shell, but also it depends upon the $\chi_{ext}$ of JMN1$_{ext}$ and the radius $r_{b2}$ where the JMN1$_{ext}$ is matched with the external Schwarzschild spacetime. Therefore, for a fix value of the radius ($r_{b1}$) of the thin matter shell, radius of the shadow can vary for different values of $r_{b2}$ and $\chi_{ext}$. On the other hand, in Schwarzschild black hole spacetime, shadow radius depends upon the Schwarzschild mass or the radius of the photon sphere only (eq.~(\ref{shadowSCH})). Therefore, shadow of the photon sphere in Schwarzschild spacetime only carries the information of Schwarzschild mass, on the other hand shadow of a thin matter shell carries the information of the radius of the shell ($r_{b1}$), $\chi_{ext}$ and the radius ($r_{b2}$) of the outer edge of the external JMN1 spacetime. So, we can say that the radius of the shadow of thin matter shell carries the information of the whole structure of spacetime.
% As we know, according to the ``No hair" theorem, a black hole can have only three properties: Mass, angular momentum and charge. Therefore, if the theorem is true, a shadow of the photon sphere of a black hole only can carry the information of the black hole mass, angular momentum and charge. On the other hand, the shadow of a thin matter shell can have the detail information of the corresponding entire structure of the spacetime. [\textcolor{green}{I think it is better to remove this ``No hair" thing.}]

Using the eq.~(\ref{spctimestruc}), one can verify that the $\chi_{ext}$ and $r_{b2}$ together fix the total Schwarzschild mass of the spacetime structure, $M_{TOT}=\frac{\chi_{ext}r_{b2}}{2}$. For the spacetime structure, in fig.~(\ref{JMNSCHshadow}), we consider the total mass ($M_{TOT}$) to be unity and we consider the mass ($M_{in}$) enclosed by the thin matter shell to be half of the total mass ($M_{TOT}$) of the entire spacetime structure. Therefore, $r_{b2}=\frac{2}{\chi_{ext}}$ and $r_{b1}=\frac{1}{\chi_{int}}$.
In fig.~(\ref{JMNSCHshadow}), we take the $\chi_{int}=0.75$ and $\chi_{ext}=0.1$ for JMN1$_{int}$ and JMN1$_{ext}$ respectively, then the radius of the thin matter shell will be $r_{b1}=1.333$ and the radius of the outer edge of the JMN1$_{ext}$ will be $r_{b2}=20$. The shadow radius in observer sky will be $b_{csp}=\frac{1.054~r_{b1}}{\left(\frac{r_{b1}}{r_{b2}}\right)^{0.055}}=1.63$, which is $1.22$ times of the radius    ($r_{b1}$) of thin matter shell. The intensity variation and the shadow of the thin matter shell for the above mentioned values of $\chi_{int}$, $\chi_{ext}$, $r_{b1}$ and $r_{b2}$, are shown in fig.~(\ref{JMNinten}), (\ref{JMNshadow}) respectively.
 
As we have mentioned, the spacetime structure given in eq.~(\ref{spctimestruc}) may be used to model the galactic halo structure, where $r_{b2}$ will be the hallo radius. In \cite{Dey:2019fja}, it is shown how this type of spacetime structure can be formed in the cosmological scenario. If one models the dark matter halo by the spacetime structure, then one has to investigate stellar motion around the galactic center. As we previously discussed, the motion of different `S' stars around the Milky-way galactic center Sgr-A* is being observed continuously by different collaborations(e.g. GRAVITY, SINFONI, etc.). As these stars are very close to the Sgr-A*, their orbit's precession due to the spacetime curvature may be observed in the near future. In \cite{Bambh}, we investigated the perihelion precession of timelike orbits in different naked singularity spacetimes and compared the results with the perihelion precession of timelike orbits in Schwarzschild spacetime. We find out that the perihelion precession in naked singularity spacetimes can be opposite to the direction of motion of particles. We showed that this opposite or negative precession cannot be possible in Schwarzschild spacetime. Therefore, any evidence of negative precession can rule out the existence of vacuum blackhole spacetime around the center of the Milky-way. As we have mentioned previously, in \cite{Bambh}, it is shown that for JMN1 naked singularity spacetime, negative precession and positive precession of timelike bound orbits are possible when $\chi<\frac13$ and $\chi>\frac13$ respectively. In fig.~(\ref{JMNSCH}), we show that both the positive and negative precession of bound timelike orbits are possible in the external JMN1 spacetime. On the other hand, we also show the shadow cast by spherically symmetric thin matter shell which exists at the matching radius of the internal and external JMN1 spacetimes.   The timelike orbit equation in external JMN1 spacetime can be written in the following form,
\begin{equation}
 \frac{d^2u}{d\phi^2} + (1 - \chi_{ext}) u - \frac{\gamma^2}{2h^2}\frac{\chi_{ext}}{(1- \chi_{ext})}\left(\frac{1}{u}\right)\left(\frac{1}{u~r_{b2}}\right)^\frac{-\chi_{ext}}{(1- \chi_{ext})}=0\,\, .
\end{equation}
Using the above orbit equation, in fig.~(\ref{JMNSCH}), we show the precession of stable timelike bound orbits in external JMN1 spacetime.
In fig.~(\ref{JMNorbitshadow1}), we show the possibility of negative precession of the timelike bound orbits and the shadow of the central thin shell of matter by considering $\chi_{int}=0.75$, $\chi_{ext}=0.3$, $r_{b2}=1000$ and $r_{b1}=\frac{1}{\chi_{int}}=1.333$. From eq.~(\ref{shadowStruc}), we get the shadow radius $b_{csp}=6.58$. As we know, for positive precession we need $\chi_{ext}>\frac13$. Therefore, in fig.~(\ref{JMNorbitshadow2}),  to get the positive precession, we consider $\chi_{ext}=0.36$. For this case, the shadow radius becomes $b_{csp}=10.7$. As we have mentioned, thin shell shadow size depends upon various parameters' values of the entire spacetime structure. Therefore, in figs.~(\ref{JMNorbitshadow1},\ref{JMNorbitshadow2}, \ref{JMNshadow}), the shadow radius of the thin shell of matter can be changed by changing different parameters' values of the proposed spacetime structure.

\section{Conclusion}
\label{conclusion}
 In this paper, we have investigated the shadow of a spherically symmetric thin matter shell and we compared that with the shadow cast by a Schwarzschild black hole. We also show that stable bound timelike orbits with positive and negative precession can be incorporated considering the spacetime structure mentioned in eq.~(\ref{spctimestruc}). To get the shadow of the thin matter shell we need $\chi_{int}>\frac23$ and $\chi_{ext}<\frac23$. On the other hand,  for positive and negative precession, we need $\chi_{ext}>\frac13$ and $\chi_{ext}<\frac13$ respectively. Few important properties of a shadow of the thin matter shell are coming out from our discussion, which are as below.  
\begin{itemize}
\item Negative precession with a central shadow is forbidden in JMN1 and JNW naked singularity spacetimes. We show that, in the proposed spacetime structure (eq.~(\ref{spctimestruc})) constructed using the JMN1$_{int}$ and JMN1$_{ext}$ naked singularity spacetimes, both shadow and negative precession can simultaneously exist. On the other hand, shadow with positive precession also can exist in the proposed spacetime configuration. Therefore, any observational result of negative precession of `S' stars with a central shadow can abandon the possibility of existence of a central black hole. In this case the proposed naked singularity spacetime configuration can be one of the candidates to explain the observed phenomena.   On the other hand, both the black hole or naked singularity can exist if  positive precession of `S' stars with central shadow is observed. One can form same type of spacetime configuration (eq.~(\ref{spctimestruc})) with JNW naked singularity spacetime which can allow negative precession of timelike bound orbits with a central shadow.
%%%%%%%%%%%%%%%%%%%%%%%%%%%%%%%%%%%%%%%%%%%%%%
\item Shadow of the thin matter shell can be formed due to the curvature around the matter shell. We do not consider any chemical absorption of light by the matter present at the thin matter shell.

\item In black hole spacetimes, the radius of the shadow only depends upon the mass, angular momentum and charge of the central object. On the other hand, the radius of the shadow of a thin matter shell carries the information of the detailed structure of internal and external spacetime (eq.~(\ref{shadowStruc})). If we model the dark matter halo structure by the structure described in eq.~(\ref{spctimestruc}), then the radius of the central shadow of the thin matter shell should be correlated with the radius of dark matter halo ($r_{b2}$) and the total dark matter halo mass ($\frac{\chi_{ext}r_{b2}}{2}$). This correlation in eq.~(\ref{shadowStruc}), can be verified when we have sufficient number of data of the central shadow radius for different galaxies. 
\end{itemize}
As was previously mentioned, in this paper, we consider Newton's gravitational constant and light velocity as unity. Therefore, in this paper, we do not attempt to fit the theoretical prediction with any observational data. Fig.~(\ref{JMNSCH}) shows only the possibility of simultaneous existence of negative precession (or positive precession) and central shadow. In that figure, we show the astrophysical importance of proposed spacetime configuration in the context of bound timelike orbits of `S' stars around the galactic center, and the shadow of the galactic center. If one wants to fit the theoretical results with `S' stars' data, one needs to consider actual values of Newton's gravitational constant and light velocity.  A detailed phenomenological discussion of `S' stars orbits and central shadow will be reported elsewhere.

%%%%%%%%%%%%%%%%%%%%%%%%%%%%%%
%%%%%% References
%%%%%%%%%%%%%%%%%%%%%%%%%%%%%%
%%%%%%%%%%%%%%%%%%%%%%%%%%%%%%
%https://cdn.journals.aps.org/files/styleguide-pr.pdf

\end{document}